	\documentclass{elsart}

	 \usepackage{graphicx}
	  \include{epsfig}
	\usepackage{amssymb}
        \usepackage{amsmath}
	\journal{Physica A}

\usepackage{longtable}
	
\begin{document}

	\begin{frontmatter}


    	\title{The Anderson-Darling test of fit for the power law distribution from left censored samples}
    	\author{H.F. Coronel-Brizio},
    	\ead{hcoronel@uv.mx}
\author{A.R. Hern\'andez-Montoya\corauthref{cor}}
\corauth[cor]{Corresponding author:
Facultad de F\'{\i}sica e Inteligencia Artificial. Departamento de Inteligencia Artificial. Universidad Veracruzana, Sebasti\'an Camacho 5, Xalapa, Veracruz 91000. M\'{e}xico}
    	\ead{alhernandez@uv.mx}
	\ead[url]{www.uv.mx/alhernandez/}

	\address{Facultad de F\'{\i}sica e Inteligencia Artificial. Departamento de Inteligencia Artificial. Universidad Veracruzana. Xalapa, Veracruz. M\'{e}xico.}

       \begin{abstract}
       \noindent
          Maximum likelihood estimation and a test of fit based on the Anderson-Darling statistic is presented for the case of the power law distribution when the parameters are estimated from a left-censored sample. Expressions for the maximum likelihood estimators and tables of asymptotic percentage points for the $A^{2}$ statistic are given. The technique is illustrated for data from the Dow Jones Industrial Average index, an example of high theoretical and practical importance in Econophysics, Finance, Physics, Biology and, in general, in other related Sciences such as Complexity Sciences.
       \end{abstract}

	\end{frontmatter}

\vspace*{-0.8cm}
\section{Introduction}

The statistical model known as the power law distribution has important applications in Natural, Social, Economical and also in Computing Sciences. Some related phenomena involving power law distribution are: scaling, universality, criticality, phase transitions \cite{scale,phase0,phase1}, fractals \cite{fractals}, complex networks \cite{nets0,nets1}, earthquakes \cite{earthquakes}, size of files in a computer system \cite{comput1}, World Wide Web Topology \cite{internet1,internet2}, Information Theory \cite{info}, financial indexes and assets price variations \cite{stanley,gopi0}, individual
income distribution \cite{pareto,gopi1}, and many more. Physicists are attempting to produce a general theory in order to explain, under an unified point of view, all this phenomena and the mechanism that drive them to produce a power law distribution. The main candidate seems to be currently known under the general name of Complex Systems  Science or Complexity Sciences. Some preliminary progress has been achieved in this direction in the context of Economical Complex Systems \cite{gab1,gab2}.

From the above said, it can be seen that the importance of the applied statistical problem of fitting a power law distribution is of great
and current interest. However, difficulties involved in performing a fit of a power law to empirical data are many and very subtle (exponent sensitivity
to a big number of data entries, to the  selection of the cut off parameter, discrimination of spurious power law distributions, etc.) and many related and very interesting papers proposing new fitting methodologies, criticizing other or proposing alternative statistical models to the power law to describe
empirical data have been recently published \cite{claus,gallegati,gold,bauke,stanley2,chin,co-arhm,scalas,iguchi,sornette,Newman}.

This paper can be considered as a proposal to formalize and establish a solid statistical procedure to perform a good power law fit in the statistical sense, by means of considering an approach based on the statistical theory of estimation and tests of fit from left censored samples. In section \ref{MLE} the problem is reviewed and formalized, introducing the required technical aspects; in section \ref{tests} the statistical fitting procedure is described; in section \ref{quad} the class of quadratic statistics and the Anderson-Darling $A^2$ statistic are defined; in section \ref{app} we present the calculations for the asymptotic percentage points used to evaluate the goodness of the fit and in section \ref{speed} results from a simulation study to investigate the speed of convergence of the calculated asymptotic percentage points to those observed empirically are shown. Section \ref{example} illustrates the fitting procedure with empirical data by means of an example from finance of a high theoretical and practical importance in Econophysics, for financial scientists and traders: a power law fit for the tails of a set of daily variations of a financial index as the Dow Jones Industrial Average (DJIA). Finally, conclusions are presented in section \ref{conclusions}.

\section{Preliminaries and Maximum likelihood estimation}
\label{MLE}
A random variable $Y$ is said to follow a power law distribution, if its cumulative distribution function (CDF) is given by

\begin{equation}
   F(y;\alpha,\theta)=1-\left(\frac{\theta}{y}\right)^{\alpha} \, \, \text{for} \, y>\theta, \, \alpha>0,\,\theta>0.
   \label{eq:F0}
\end{equation}

and it is required to test goodness of fit to the tails of this distribution using the largest values from the sample of their returns. This situation can be viewed performing estimation and goodness of fit under type II left censoring, which corresponds to the case in which, for fixed $r$, the $n-r$ smallest observations are missing, so the estimation and test procedures are based on the largest $r$ sample values $y_{(n-r+1)},y_{(n-r+2)},\ldots,y_{(n)}$ where $y_{(i)}$ denotes the $i-$th order statistic in a sample of size $n$ from an absolutely continuous distribution $F$.

Given that the Anderson-Darling $A^{2}$ is known to be a powerful statistic for detecting departures in the tails from the hypothesized distribution, it becomes necessary to obtain its asymptotic distribution when the parameters of the distribution have been estimated based on a left censored sample. Details on maximum likelihood estimation for the case of complete samples for the Power law distribution can be found, for instance, in \cite{gab2} and \cite{Newman}, appendix B.

The log-likelihood for a left-censored sample $y_{(n-r+1)},\ldots,y_{(n)}$ from the distribution \eqref{eq:F0}, is given by
\[
   l(\alpha,\theta)=(n-r) \ln\!\left[ 1-\!\left(\frac{\theta}{y_{(n-r+1)}}\right)^{\alpha} \right] +r\ln(\alpha) +r\alpha\, \ln(\theta) - \left( \alpha+1 \right) \sum_{i=1}^{r} \ln y_{(n-r+i)} 
\]
The maximum likelihood estimators (MLE) of the parameters $\alpha$ and $\theta$, are the solution of the equations
\begin{eqnarray}
    \frac{\partial l(\alpha,\theta)}{\partial \alpha}&=&{\frac {r}{\alpha}}+r\ln  \left( \theta \right) -\sum _{i=1}^{r}\ln y_{{(n-r+i)}}  \nonumber \\ 
     && - \left( n-r \right)  \left( {\frac {\theta}{y_{{(n-r+1)}}}} \right) ^{\alpha}\ln  \left( {\frac {\theta}{y_{{(n-r+1)}}}} \right)  \left[ 1-\left( {\frac {\theta}{y_{{(n-r+1)}}}} \right) ^{\alpha} \right] ^{-1} \nonumber \\
      &=& 0 \label{eq:eqalpha} \\
    \frac{\partial l(\alpha,\theta)}{\partial \theta}&=&- \left( n-r \right)  \left( {\frac {\theta}{y_{{(n-r+1)}}}} \right) ^{\alpha}\alpha{\theta}^{-1} \left[ 1- \left( {\frac {\theta}{y_{{(n-r+1)} }}} \right) ^{\alpha} \right] ^{-1}+{\frac {r\alpha}{\theta}}  \nonumber \\
      &=&0 \label{eq:eqtheta}
\end{eqnarray}
It is easy to verify that
\begin{eqnarray}
    \qquad\quad \hat{\alpha}&=&r \left[{\sum _{i=1}^{r}\ln 
      y_{{(n-r+i)}} -r\ln  y_{{(n-r+1)}}} \right]^{-1} \label{eq:mlealpha} \\
    \hat{\theta}&=&\left(r/n\right)^{1/\hat{\alpha}}y_{(n-r+1)} \label{eq:mletheta}
\end{eqnarray}
Let us denote Fisher's information matrix by
\[
   I(\alpha,\theta)=- E\left[ 
   \begin {array}{cc} 
      {\frac {\partial ^{2}}{\partial {\alpha}^{2}}}l \left( \alpha,\theta \right) &{\frac {\partial ^{2}}{\partial \alpha\partial \theta}}l \left( \alpha,\theta \right) \\
      \noalign{\medskip}{\frac {\partial ^{2}}{\partial \alpha\partial \theta}} l \left( \alpha,\theta \right) &{\frac {\partial ^{2}}{\partial {\theta}^{2}}}l \left( \alpha,\theta \right)
   \end {array}
  \right] 
\]
Asumming that the proportion of censoring $q=1-r/n$  remains constant as $n\rightarrow \infty$, we obtain the following limiting expressions:
\begin{equation*}
   V=\lim_{n\rightarrow \infty} \frac{1}{n}I(\alpha,\theta)= \left[ 
   \begin {array}{cc}
      {\frac { \left( 1-q \right)  \left[ q+ \left( \ln  \left( 1-q \right)  \right) ^{2} \right] }{{\alpha}^{2}q}}&{\frac { \left( 1-q \right) \ln  \left( 1-q \right) }{\theta\,q}} \\
      \noalign{\medskip} {\frac { \left( 1-q \right) \ln  \left( 1-q
    \right) }{\theta\,q}} & {\frac {{\alpha}^{2} \left( 1-q \right) }{q{
   \theta}^{2}}}
   \end {array}
   \right] 
\end{equation*}
Thus, the asymptotic variance-covariance matrix of the maximum likelihood estimators will assume the form $n^{-1}V^{-1}$, where
\[
  V^{-1}=  \left[ 
   \begin {array}{cc}
      {\frac {{\alpha}^{2}}{1-q}}&{-\frac {\ln \left( 1-q \right) \theta}{1-q}} \\
      \noalign{\medskip}
      {-\frac {\ln \left( 1-q \right) \theta}{1-q}} & {\frac { \left[ q+ \left( \ln \left( 1-q \right)  \right) ^{2} \right] {\theta}^{2}}{{\alpha}^{2} \left( 1-q \right) }}
   \end {array}
   \right] 
\]
For the case of complete samples (i.e. $q=0$) the estimate $\hat{\theta}=y_{(1)}$ is super-efficient in the sense that its asymptotic variance is $O(n^{-2}),$ so $n^{-1}\lim_{q \rightarrow 0} \text{Var}(\hat{\theta})$ $=0$; in fact, for $n>2$, $\text{Var}(\hat{\theta})={\theta}^{2}n\alpha\left[ \left(\alpha\,n -2 \right)  \left(\alpha\,n-1 \right) ^{2} \right]^{-1}$. For practical applications this means that the asymptotic distributional results will be identical to the case when the parameter $\theta$ is known. On the other hand, $ \text{Var}(\hat{\alpha}) = n^{-1}\lim_{q\rightarrow 0} \alpha^{2}/(1-q) = \alpha^{2}/n $.  

\section{Test procedures}
\label{tests}

Suppose that we are interested in testing the null hypothesis that the random sample $y_{1},\ldots,y_n$, was drawn from the distribution \eqref{eq:F0}, based on the $r$ largest observations. The test can be performed as follows:
\begin{enumerate}
   \item Find the maximum likelihood estimators $\hat{\alpha}$ and $\hat{\theta}$ of the parameters $\alpha$ and $\theta$ in \eqref{eq:F0}, using formulas \eqref{eq:mlealpha} and \eqref{eq:mletheta}.
   \item Obtain the order statistics $y_{(n-r+1)}\leq \ldots \leq y_{(n)}$ and compute $z_{(n-i+1)}=F\left[y_{(n-i+1)};\hat{\alpha},\hat{\theta}\right]$ for $i=1,\ldots,r$. 
   \item Compute the Anderson-Darling in its version for a type II left-censored sample
      \begin{eqnarray*}
         \!\!\!\!\!\!\!\!\!\! A^{2}_{n,r} &=& -\frac{1}{n} \sum _{i=1}^{r} \left( 2\,i-1 \right)  \left\{ \ln  \left[ 1-z_{(n-i+1)} \right] -\ln z_{(n-i+1)} \right \} - 2\,\sum _{i=1}^{r}\ln  z_{(n-i+1)}  \\
           && - \frac{1}{n} \left[ \left( r-n \right)^{2}\ln z_{(n-r+1)} -{r}^{2}\ln  \left( 1-z_{(n-r+1)} \right) + n^{2} \left( 1-z_{(n-r+1)} \right) \right] 
      \end{eqnarray*}
  \item Using the value $q=1-r/n$, the proportion of left-censoring, refer to table \ref{tab:tA2}. If the value of the test statistic exceeds the value in the table, for a given significance level, reject the null hypothesis.
\end{enumerate}

\begin{table}[!htp]
  \caption{Upper percentage points of the asymptotic distribution of the $A^2_{n,r}$ statistic, for selected censoring proportions.}
  \label{tab:tA2}
  \begin{center}
  \begin{tabular}{|c|c|c|c|c|c|c|c|}
     \multicolumn{1}{c}{Censoring proportion } & \multicolumn{5}{c}{Significance level} \\
     \hline
     $q=1-r/n$& \textbf{0.15}&\textbf{0.10}&\textbf{0.05}&\textbf{0.025}&\textbf{0.01} \\
     \hline
     \hline
      0.00&  0.9123& 1.0588& 1.3181& 1.5873& 1.9554\\
      0.05&  0.7364& 0.8566& 1.0695& 1.2905& 1.5925\\
      0.10&  0.6354& 0.7388& 0.9217& 1.1114& 1.3706\\
      0.15&  0.5584& 0.6489& 0.8087& 0.9743& 1.2005\\
      0.20&  0.4950& 0.5748& 0.7157& 0.8616& 1.0607\\
      0.25&  0.4406& 0.5114& 0.6361& 0.7652& 0.9414\\
      0.30&  0.3928& 0.4557& 0.5663& 0.6808& 0.8368\\
      0.35&  0.3500& 0.4058& 0.5039& 0.6054& 0.7436\\
      0.40&  0.3111& 0.3606& 0.4474& 0.5372& 0.6594\\
      0.45&  0.2755& 0.3191& 0.3957& 0.4748& 0.5825\\
      0.50&  0.2425& 0.2808& 0.3480& 0.4173& 0.5117\\
      0.55&  0.2118& 0.2451& 0.3036& 0.3639& 0.4460\\
      0.60&  0.1830& 0.2118& 0.2621& 0.3141& 0.3847\\
      0.65&  0.1559& 0.1804& 0.2232& 0.2673& 0.3273\\
      0.70&  0.1303& 0.1507& 0.1864& 0.2231& 0.2731\\
      0.75&  0.1060& 0.1226& 0.1515& 0.1813& 0.2219\\
      0.80&  0.0829& 0.0958& 0.1184& 0.1417& 0.1732\\
      0.85&  0.0608& 0.0703& 0.0868& 0.1039& 0.1270\\
      0.90&  0.0397& 0.0459& 0.0567& 0.0677& 0.0828\\
      0.95&  0.0195& 0.0225& 0.0278& 0.0332& 0.0405\\
     \hline
  \end{tabular}
  \end{center}
\end{table}

\section{Quadratic statistics and asymptotic theory}
\label{quad}
The Anderson-Darling $A^{2}$ statistics belongs to a class of discrepancy measures of the form
\[ Q_{n}=n\int_{-\infty}^{\infty}\left[F_{n}(x)-F(x;\boldsymbol{\theta})\right]^{2} \psi(x) dF(x;\boldsymbol{\theta}) \]
known as \emph{quadratic statistics}, where $F_{n}$ denotes the Empirical Distribution Function (EDF) of a random sample $x_{1},\ldots,x_{n}$ from an absolutely continuous distribution $F$, $\boldsymbol{\theta}$ denotes a vector parameter and $\psi$ is a weighting function.

When $\psi(x)=1$, the resulting statistic is the well known Cram\'{e}r-von Mises $W^2$. In order to put more
weight in the tails, the Anderson-Darling $A^2$ statistic is obtained for $\psi(x)=\left\{\left[F(x;\boldsymbol{\theta})\right]\left[1-F(x;\boldsymbol{\theta})\right]\right\}^{-1}$.

In this section, the process of finding the asymptotic distribution of the EDF statistics, is briefly
summarized. For a more detailed treatment of the subject, the reader is referred to the works by D'Agostino, Durbin and Stephens \cite{Dagostino,Durbin73,Stephens76}.

The asymptotic theory for doubly censored samples with known parameters, has been given in Pettitt and Stephens
\cite{PettittAndStephens76}. Pettitt \cite{Pettitt76} modified the theory in Durbin \cite{Durbin73} for testing normality from censored samples with parameters estimated by maximum likelihood. Here, these results are used to obtain the asymptotic distribution for the power law distribution under type II censoring. In the following, it will be assumed that the proportion censored, $q=1-r/n$, remains constant as $n$ tends to infinity. 

Let $\hat{\boldsymbol{\theta}}$ denote the maximum likelihood estimator of the vector parameter $\boldsymbol{\theta}$, with estimates where necessary. For a singly left-censored sample, the process $\sqrt{n}\left\{F_n(x)-F(x;\hat{\boldsymbol{\theta}})\right\}$, evaluated at $t=F(x;\hat{\boldsymbol{\theta}})$, converges weakly to a Gaussian process $\left\{Y(t):t \in (q,1) \right\}$ with certain covariance function $\rho(s,t)$. The limiting distribution will depend on the functional form of $F$, and on which parameters are being estimated. 

The statistic $A^2$ is asymptotically a functional of the process $Y(t)$; namely,  $A^{2}$ converges weakly to $\displaystyle\int_{q}^{1}a^{2}(t)dt$ where $a(t)=Y(t)\left[\sqrt{t(1-t)}\right]^{-1}$. $Y(t)$ and $a(t)$ are both Gaussian processes defined in $(q,1)$, with covariance functions $\rho(s,t)$ and $\rho_a(s,t)=\rho(s,t)\left[(s-s^2)(t-t^2)\right]^{-1}$, respectively, for $q\leq s,t \leq 1$.

It is known that the limiting distribution (see for example, Durbin \cite{Durbin73}) is that of $\sum_{i=1}^{\infty} \lambda_{i}\nu_{i}$, where $\nu_{1},\ldots$ are independent chi-square random variables with one degree of freedom, and $\lambda_{1}^{*},\ldots$ are the eigenvalues of the integral equation
\begin{equation}
   \int_{q}^{1}\rho^{*}(s,t)f_i(s)ds=\lambda_i^* f_i(t)
   \label{eq:inteq}
\end{equation}
where $\rho^{*}$ denotes the covariance function corresponding to the limiting process on which the test statistic is based; in our case, $\rho_a(s,t)$.

\section{Asymptotic percentage points}
\label{app}

In samples from the power law distribution defined in \eqref{eq:F0}, with a left-censored proportion $q$, the limiting Gaussian process was found to have a covariance function given by
\begin{eqnarray*}
   \rho(s,t) &=& \min(s,t)-st-\left( 1-s \right)  \left( 1-t \right) (1-q)^{-1} \left[ \ln  \left( 1-s \right) \ln  \left( 1-t \right) \right. \\
    && - \left.\ln  \left( 1-t \right) \ln \left( 1-q \right) -\ln  \left( 1-s \right) \ln  \left( 1-q \right) 
   + q +  \ln^{2} \left( 1-q \right)  \right]
\end{eqnarray*}
Therefore the asymptotic distribution of the Anderson-Darling statistic will not depend on the particular values of the parameters $\theta$ and $\alpha$.

Also, for $q=0$, when the full sample is available,
\[
   \rho(s,t)=min(s,t)-st-(1-s)(1-t)\ln(1-s)\ln(1-t)
\]
The asymptotic points were found numerically using 400 points in $(q,1)$ to approximate the integral and solve \eqref{eq:inteq}. In a $400\times 400$ grid, the appropriate covariance function was evaluated and the eigenvalue problem solved for different values of $q$, the proportion of censoring, ranging from 0.05 to 0.95, with increments of 0.05 units. These eigenvalues were then used to calculate the asymptotic percentage points using Imhof's method \cite{Imhof61}. The results are shown in table \ref{tab:tA2}. The row corresponding to $q=0$ denotes the asymptotic percentage points for complete samples.

\section{Small sample distributions}
\label{speed}
A simulation study was performed to investigate the speed of convergence of the empirical percentage points to the asymptotic ones, for both statistics and considering different censoring proportions. For each combination of $q$ and $n$, ten thousand pseudo-random samples from the distribution \eqref{eq:F0} were generated and the statistic $A_{r,n}^{2}$ was then calculated to estimate the empirical percentage points. The results are shown in table \ref{tab:tA2rn}. The standard errors of the estimated percentage points, say $\xi_{1-\alpha}$, for a given significance level $\alpha$, were approximated using the asymptotic expression $\hat{SE}(\xi_{1-\alpha})=\frac{1}{g(\xi_{1-\alpha})}\sqrt{\frac{\alpha(1-\alpha)}{N}}$, where $N$ denotes the number of simulations (in this case, $N=10000$) and $g$ is the density of the simulated test statistic. The density at each point can be estimated by approximating the derivative of the empirical cumulative distribution function using two adjacent percentage points.

{\small
\begin{longtable}{|c|c|c|c|c|c|}
   \caption{Empirical percentage points of the statistic $A^2_{n,r}$ for selected censoring proportions and sample sizes. The estimated standard errors are shown within parenthesis.}
\label{tab:tA2rn} \\

\multicolumn{1}{c}{Censoring}  & \multicolumn{5}{c}{Significance} \\[-.4cm]
\multicolumn{1}{c}{proportion}  & \multicolumn{5}{c}{level} \\
\hline \multicolumn{1}{|c|}{$q=1-r/n$} & \multicolumn{1}{c|}{$n$} & \multicolumn{1}{c|}{\textbf{0.15}} & \multicolumn{1}{c|}{\textbf{0.10}} & \multicolumn{1}{c|}{\textbf{0.05}} & \multicolumn{1}{c|}{\textbf{0.01}} \\ \hline 
\endfirsthead

\multicolumn{6}{c}{{\bfseries \tablename\ \thetable{} -- continuation}} \\
\hline \multicolumn{1}{|c|}{$q=1-r/n$} & \multicolumn{1}{c|}{$n$} & \multicolumn{1}{c|}{\textbf{0.15}} & \multicolumn{1}{c|}{\textbf{0.10}} & \multicolumn{1}{c|}{\textbf{0.05}} & \multicolumn{1}{c|}{\textbf{0.01}} \\ \hline 
\endhead

\hline
\endfoot

\hline
\endlastfoot

 0.050& 100&  0.7237 &0.8464 &1.0406 &1.5848 \\
      &    &  (0.88$\times 10^{-2}$) &(0.12$\times 10^{-1}$) &(0.30$\times 10^{-1}$) &(0.14$\times 10^{-1}$) \\
      & 300&  0.7318 &0.8434 &1.0450 &1.5880 \\
      &    &  (0.80$\times 10^{-2}$) &(0.12$\times 10^{-1}$) &(0.30$\times 10^{-1}$) &(0.14$\times 10^{-1}$) \\
      &$\infty$&0.7364& 0.8566& 1.0695& 1.5925 \\
 \hline
 0.100& 100&  0.6485 &0.7525 &0.9404 &1.3762 \\
      &    &  (0.74$\times 10^{-2}$) &(0.11$\times 10^{-1}$) &(0.24$\times 10^{-1}$) &(0.11$\times 10^{-1}$) \\
      & 300&  0.6269 &0.7244 &0.9053 &1.3509 \\
      &    &  (0.70$\times 10^{-2}$) &(0.11$\times 10^{-1}$) &(0.24$\times 10^{-1}$) &(0.11$\times 10^{-1}$) \\
      &$\infty$&0.6354& 0.7388& 0.9217& 1.3706\\
\hline
 0.250& 100&  0.4582 &0.5292 &0.6541 &0.9693  \\
      &    &  (0.51$\times 10^{-2}$) & (0.75$\times 10^{-2}$) & (0.17$\times 10^{-1}$) & (0.78$\times 10^{-2}$)  \\
      & 300&  0.4419 &0.5129 &0.6400 &0.9568  \\
      &  &  (0.51$\times 10^{-2}$) & (0.76$\times 10^{-2}$) & (0.17$\times 10^{-1}$) & (0.79$\times 10^{-2}$)  \\ 
      &$\infty$&0.4406&0.5114& 0.6361& 0.9414\\
 \hline
 0.500& 100&  0.2470 &0.2840 &0.3550 &0.5184  \\
     &     &  (0.26$\times 10^{-2}$) & (0.43$\times 10^{-2}$) & (0.89$\times 10^{-2}$) & (0.41$\times 10^{-2}$)  \\
     &  300&  0.2459 &0.2838 &0.3526 &0.5080 \\
     &     &  (0.27$\times 10^{-2}$) & (0.41$\times 10^{-2}$) & (0.85$\times 10^{-2}$) & (0.39$\times 10^{-2}$)  \\
          & $\infty$ &0.2425& 0.2808& 0.3480&0.5117 \\
 \hline
0.750&  100&  0.1052 &0.1209 &0.1516 &0.2247  \\
       &   &  (0.11$\times 10^{-2}$) & (0.18$\times 10^{-2}$) & (0.40$\times 10^{-2}$) & (0.18$\times 10^{-2}$)  \\
     & 300&  0.1058 &0.1238 &0.1517 &0.2168  \\
     &    &  (0.13$\times 10^{-2}$) & (0.17$\times 10^{-2}$) & (0.35$\times 10^{-2}$) & (0.16$\times 10^{-2}$)  \\        
     & $\infty$&  0.1060& 0.1226& 0.1515 &0.2219\\
\hline
0.900& 100&  0.0360 &0.0413 &0.0506 &0.0749 \\
     &    &  (0.38$\times 10^{-3}$) & (0.56$\times 10^{-3}$) & (0.13$\times 10^{-2}$) & (0.60$\times 10^{-3}$)  \\
     & 300&  0.0388 &0.0450 &0.0549 &0.0804  \\
     &    &  (0.44$\times 10^{-3}$) & (0.59$\times 10^{-3}$) & (0.14$\times 10^{-2}$) & (0.63$\times 10^{-3}$)  \\
     & $\infty$&   0.0397& 0.0459 &0.0567& 0.0828\\
 \hline
 0.950& 100&  0.0199 &0.0226 &0.0273 &0.0395  \\
      &    &  (0.19$\times 10^{-3}$) & (0.28$\times 10^{-3}$) & (0.66$\times 10^{-3}$) & (0.30$\times 10^{-3}$)  \\
      & 300&  0.0197 &0.0228 &0.0282 &0.0422  \\
      &    &  (0.22$\times 10^{-3}$) & (0.32$\times 10^{-3}$) & (0.76$\times 10^{-3}$) & (0.35$\times 10^{-3}$)  \\
      &$\infty$& 0.0195& 0.0225& 0.0278& 0.0405 \\
\end{longtable}
}

These results seem to suggest that the speed of convergence of the empirical, to the asymptotic percentage points, does not depend on the proportion of censoring, at least in a significant way. Thus, the asymptotic percentage points can be used with good accuracy for moderately large $n$. 

\section{An example taken from Finance: Fitting the tails of the Dow Jones Index Daily Variations}
\label{example}
\begin{figure}[!htb]
   \begin{center}
   \includegraphics[width=0.65\textwidth,bb=4 4 520 350,clip]{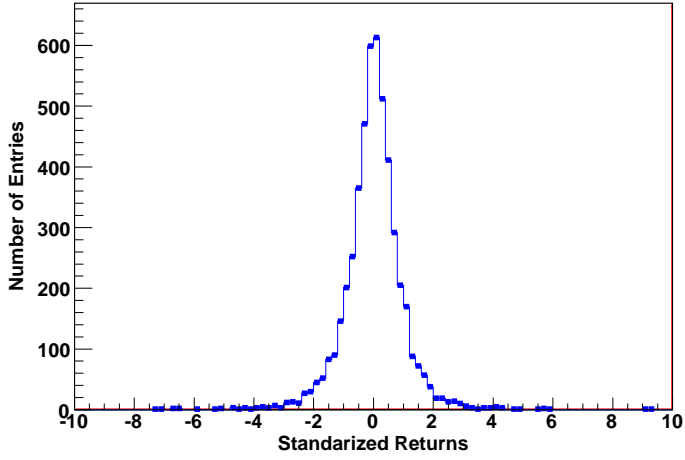}
   	\caption{Histogram for standardized returns. It has 5001 entries.}
   	\label{fig:histo1}
   \end{center}
\end{figure}

In order to illustrate the technique\footnote{A program to perform the analysis described here, is available on  request to  hcoronel@uv.mx. Later upgraded  versions, will be available in a more formal web site.}, we consider the series consisting of 5001 standardized returns computed from the daily closing values of the Dow Jones Industrial Average Index (DJIA). The data includes values from January 1st, 1990 to November 3rd, 2009 and its file can be obtained in www.yahoo.com. The histogram for the set of standardized returns is shown in figure \ref{fig:histo1}.

\begin{figure}[!htb]
\begin{center}
	\includegraphics[width=0.65\textwidth]{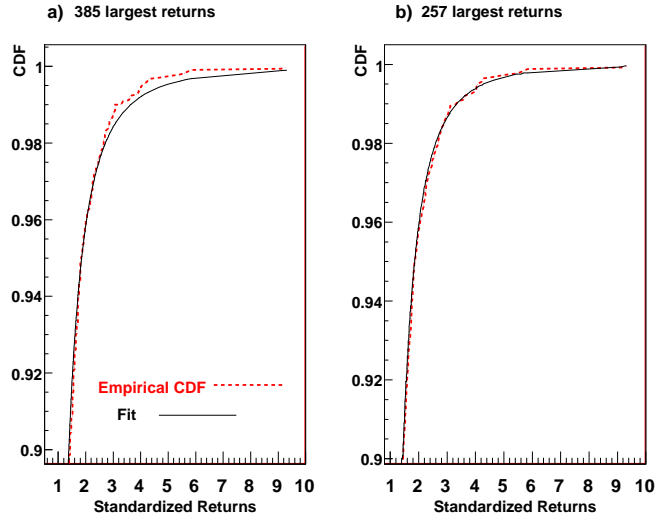}
	\caption{Fitted (solid line) and empirical (dashed line) cumulative distribution functions
	(CDF) for: a)  The largest 385 positive returns. b) The largest 257 positive returns.}
	\label{fig:fitpos}
\end{center}
\end{figure}

Suppose that we are interested in performing a power law fit based on the largest $r=385$ positive values of the standardized returns. Using formulas \eqref{eq:mletheta} and \eqref{eq:mlealpha} we obtain $\hat{\theta}=0.528$ and $\hat{\alpha}=2.387$. The calculated value of the Anderson-Darling statistic is $A^{2}_{n,r}=0.18$ which exceeds the value in table \ref{tab:tA2} for a proportion of censoring $q=0.85$ and significance level $0.01$. It is then concluded that the power law model should be rejected with a probability value $p<0.01$. See figure \ref{fig:fitpos}a. 

A second fit based on the $r=257$ largest positive returns, gives $\hat{\theta}=0.622$, $\hat{\alpha}=2.728$ and $A^{2}_{n,r}=0.027$ which is less than the $0.15$ critical point for a censoring rate $q=0.90$. In this case the power law fit is not rejected with an associated probability value $p>0.15$, so the fit is considered good. Figure \ref{fig:fitpos}b shows the resulting fit.

\begin{figure}[!htb]
\begin{center}
	\includegraphics[width=0.65\textwidth]{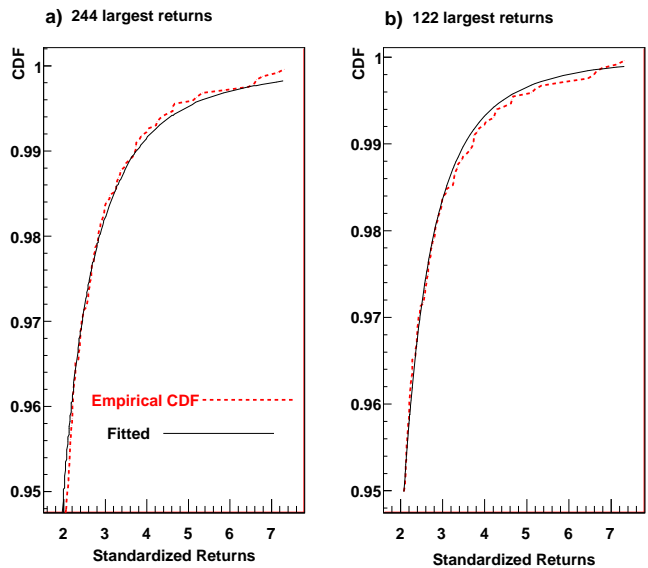}
	\caption{Fitted (solid line) and empirical (dashed line) cumulative distribution functions (CDF) 
	for: a) The largest 244 absolute negative returns. b) The largest 122
	absolute negative returns.}
	\label{fig:fitneg}
\end{center}
\end{figure}

This procedure can be applied to the left tail, considering  the absolute values of the $2436$ negative returns. A test based on the largest $r=244$ values, corresponding to a censoring proportion $q=0.90$, gives $\hat{\theta}=0.625$, $\hat{\alpha}=2.562$. The value $A^{2}_{n,r}=0.077$ gives a probability value $0.01<p<0.025$, indicating strong evidence against the power law fit shown in figure \ref{fig:fitneg}a. If we now consider a censoring proportion $q=0.95$, the results for $r=122$ are $\hat{\theta}=0.785$, $\hat{\alpha}=3.066$ and $A^{2}_{n,r}=0.009$; the associated probability value is now $p>0.15$, indicating a very good fit. This fit is shown in figure \ref{fig:fitneg}b.

\section{Conclusions}
\label{conclusions}
Applying the theory in Durbin \cite{Durbin73}, the percentage points of the asymptotic distribution of the Anderson-Darling $A^{2}$ statistic were obtained numerically and tables for testing goodness of fit for the power law distribution, when the parameters are estimated from a left-censored sample, were provided. Results from a simulation study showed that a test of fit for this distribution can be performed with good accuracy using the asymptotic percentage points for moderately large samples. It was also found that the speed of convergence of the empirical to the asymptotic percentage points, does not show a significant dependence on the censoring rate.

Given that the test is based on the Anderson-Darling $A^{2}$ statistic, which puts more weight in the tails of the distribution, the resulting test appears to be demanding as it can be concluded from the cases described in the example.

\vspace{0.5cm}
{\bf Acknowledgments\\}
\noindent
The authors acknowledge support from the Sistema Nacional de Investigadores (SNI-CONACyT, M\'exico) and support also from CONACyT under project number 129141.

\end{document}